%Paper: hep-th/9205012
%From: grisaru@vxcrna.cern.ch
%Date: Thu, 7 May 92 11:17:51 +0200

\magnification\magstep1
\tolerance=1600
\parskip=0pt
\baselineskip= 6 true mm

\def\di#1#2{{\rm d}^{#1}\! {#2}}
\def\ft#1#2{{\textstyle{{#1}\over{#2}}}}
\def\a{\alpha}
\def\b{\beta}
\def\g{\gamma}\def\G{\Gamma}
\def\d{\delta}
\def\e{\epsilon}

\def\m{\mu}

\def\n{\nu}
\def\p{\psi}\def\P{\Psi}
\def\r{\rho}
\def\s{\sigma}

\def\x{\chi}
\def\o{\omega}
\def\ve{\varphi}

\def\slash{\llap /}
\def\lagr{{\cal L}}
\def\pa{\partial}

{\nopagenumbers

\vglue 1truecm
\rightline{CERN-TH.6477/92}
\rightline{THU-92/04}
\rightline{hepth@xxx/9205012}
\rightline{April 1992}

\vskip 1.5truecm
\centerline{{\bf TWO-LOOP FINITENESS OF $D=2$ SUPERGRAVITY }}

\vskip 2truecm

\centerline{B.~de Wit\footnote{${}^1$}{Permanent address:
Institute for Theoretical Physics, University of
Utrecht,\hfil\break\indent The Netherlands.},
M.T.~Grisaru\footnote{${}^2$}{Permanent address: Brandeis
University, Waltham, MA 02254, USA.},
E.~Rabinovici\footnote{${}^3$}{Permanent address: Racah Institute
of Physics, Jerusalem, Israel. \hfil \break \indent
This work is supported in part by the Israel Academy of Science
and by the \hfil\break\indent
BSF-American-Israeli Bi-National Science Foundation. } }
\vskip .1truecm
\centerline{Theory Division, CERN}
\centerline{CH-1211, Geneva 23, Switzerland}
\vskip .5truecm
\centerline{and}
\vskip .5truecm
\centerline{H.~Nicolai}
\vskip .1truecm
\centerline{II. Institute for Theoretical Physics, University of
Hamburg}
\centerline{Luruper Chaussee 149, 2000 Hamburg 50, F.R.G. }
\vskip 2true cm

\centerline{\bf Abstract}
\vskip .2truecm
{\narrower\noindent We establish two-loop (on shell) finiteness
of certain supergravity theories in two dimensions. Possible
implications of this result are discussed.\smallskip}

\footnote{}{}
\footnote{}{CERN-TH.6477/92}
\footnote{}{THU-92/04}
\footnote{}{April 1992}

\vfill\eject
}
\pageno=1

In this paper we study the quantum properties of two-dimensional
systems of supergravity coupled to matter
that emerge by dimensional reduction from pure four-dimensional
supergravity (although our results are somewhat more general as
we shall indicate later). As any two-dimensional theory of gravity that
is free of ultraviolet divergences can be utilized to define a
related critical theory (i.e., a theory with vanishing total
central charge),
our hope is that the improved short-distance properties
of extended supersymmetric theories will enable us to establish the
existence of new consistent theories of quantum gravity.
The main result of this paper is that there is a class of
extended supergravity theories that is at least two-loop finite,
so that this hope is at least partially realized.

So far it has been shown that the only consistent critical
bosonic theories
are based either on at least twenty six (matter) fields
or at most two  [1] (for
reviews, see [2]). The latter lead to systems with
at most zero (propagating) degrees of freedom, so that the theories
tend to be over-constrained when not treated as topological
theories (for a discussion, see
[3]). In the search for critical systems with a richer
structure one
may choose to study dimensionally reduced versions of
four-dimensional general relativity. In the case of pure gravity
this proved a fruitful approach, which leads to a well-defined
theory of topological nature (for a review of topological
theories, see [4]) that governs the constant-curvature
solutions and is related to classical Liouville theory. This
approach avoids a negative number of (propagating) degrees of
freedom, and has for instance been advocated in [5,6].
The two-dimensional analogue of the Einstein equation
is imposed by a Lagrange multiplier field, which is either
introduced by hand or arises naturally from the
four-dimensional theory by standard dimensional reduction.
(Note that in the latter case this field is restricted to
be positive, so it is not a true multiplier field). The
resulting theory can also be cast in more geometrical form
[7].

Classically  these theories are not invariant
under Weyl rescalings of the two-dimensional metric.
Here we follow the standard approach and extract a scale
factor from the metric (sometimes called a compensating field),
$$
g_{\m\n} ={\rm e}^{2\s}\, \hat g_{\m\n}\, , \eqno(1)
$$
which is included into the dynamics, keeping the reference metric
$\hat g_{\m\n}$ fixed.
Of course, as the decomposition (1) is
determined up to an arbitrary factor (i.e., it is invariant
under simultaneous rescalings of $\exp (-2\s)$ and $\hat g$),
the theories are now formally
invariant under Weyl rescalings of the metric $\hat g$ and have a
traceless stress tensor. Thus they take the form of a
conformal field theory in a background metric $\hat g$. This
approach can be applied to any two-dimensional generally
covariant field theory. However, the relevant question is
whether this defines a consistent critical model. In order for
this to be the case a minimal condition is that the theory is
free of (on-shell) ultraviolet divergences
(for a discussion of the quantum aspects of compensating fields,
see [8]).

The models that we consider originate from four-dimensional
supergravity by straightforward reduction to two dimensions, so
that the fields of the four-dimensional theory depend only
on two coordinates.\footnote{${}^1$}{
The dimensional reduction at first sight appears to be
different from the more familiar ``spherical truncation", where
one compactifies the theory on $S^2$ and suppresses all
dependence on the angular coordinates (this leads for instance to
an effective theory for the radial modes of black-hole
solutions; for recent applications, see e.g. [9]). Nevertheless
it is known that the ``naive" dimensional reduction
of Einstein's theory reproduces not only the Schwarzschild solution
but many other solutions as well (stationary axisymmetric and
colliding plane-wave solutions). For the stationary axisymmetric
solutions, this reduction directly leads to the so-called
``Weyl canonical coordinates", where the dilaton field $\rho$ (see
below) is identified with a cylindrical coordinate after fixing the
residual conformal gauge invariance. For a discussion of these
and related issues, we refer to [10-12].}
Many of the features we find are thus consequences of
the higher-dimensional theory.  An important reason for studying
the quantum properties of the two-dimensional theories is their
intriguing symmetry structure,
which remains somewhat mysterious even at the classical level.
For pure gravity, it has been known for a long time that there is an
infinite-dimensional symmetry group [13]
acting on the space of solutions of Einstein's equations with two
(commuting) Killing vectors (see [10] for a review). The connection
between this group and the so-called ``hidden symmetries"
of extended supergravity theories was first emphasized and studied
in [14] and subsequently elaborated in [11,15,12].
All the models obtained by dimensional reduction of gravity and
supergravity to two dimensions are classically integrable in the
sense that they admit linear systems (or Lax pairs) for their
non-linear field equations [16,11,15]. Through this work
it has been established that the emergence of infinite dimensional
symmetries of the Kac-Moody type, which are realized by non-linear
and non-local transformations and which generalize
the corresponding (finite-dimensional) symmetries of non-linear
sigma models in higher dimensions, is a generic phenomenon in the
reduction to two dimensions. Furthermore, the $G/H$ coset structure
present in higher-dimensional supergravity theories has a natural
analogue in two dimensions, inasmuch as the (bosonic) ``manifold
of solutions" can be understood in terms of the
infinite-dimensional coset space
$G^\infty / H^\infty$, where $G^\infty$ denotes the (centrally
extended) affine extension of $G$,
and $H^\infty$ its maximally
compact subgroup with respect to the generalized Cartan-Killing form
on the Kac-Moody algebra of $G$. Experience with flat-space
integrable models
suggests that these symmetries will be of prime importance
for the quantized theories, perhaps leading to examples of
quantum integrable models of (super)gravity.

In order to introduce the models it is convenient to consider
three-dimensional supergravity at an intermediate stage of the
dimensional reduction,
$$
\lagr = \ft12 \e^{mnp}\,\Big\{\ft12 e_m^{\;a} \, R_{np a}(\o) +
\bar\p^I_m
D_n(\o) \p^I_p\Big\} + e \lagr^{\rm matter} \,, \eqno(2)
$$
where $m,n,p,\ldots$ and $a,b,\ldots$ denote three-dimensional
world and tangent space indices, respectively.
The basic supergravity fields are the dreibein
$e_m^{\,a}$, the spin-connection
field $\o_{ma}$ (with corresponding curvature
$R_{mna}(\o)$) and $N$ (Majorana) gravitino fields $\p^I_m$.
The first two terms describe three-dimensional pure $N$-extended
supergravity, which is locally supersymmetric irrespective of
the value for $N$, the number of independent supersymmetries. The
graviton and gravitino fields do not correspond to (propagating)
physical degrees of freedom, and without
the last term the theory is topological [17].
The matter Lagrangian takes the form of a supersymmetric sigma model,
coupled to the supergravitional fields. As we concentrate on
theories that originate from four-dimensional pure supergravity,
the sigma model has a homogeneous symmetric target-space metric.
We note that only a few
three-dimensional theories have been constructed explicitly so far,
and that our results hinge on certain plausible assumptions
as far as those models are concerned that have not been constructed
explicitly. The $N=16$ theory with target space
$E_{8(+8)}/SO(16)$ and a class of $N=8$ theories
based on the coset spaces $SO(8,n)/(SO(8) \otimes SO(n))$
have been given in [18]; the simpler $N=2$ theory has been
discussed in [12]. The structure of some of the other theories
can be deduced in principle from the corresponding four-dimensional
theories or by truncation of the $N=16$ theory.

The matter Lagrangians that we consider are based on homogeneous
spaces $G/H$, where $G$ is non-compact and $H$ its maximally compact
subgroup. For reasons of supersymmetry the
isotropy subgroup $H$ has the direct
product form $H=SO(N)\otimes H'$ (the subgroup $H'$ is associated with
the centralizer of the $SO(N)$ Clifford algebra in
the real representation and may be trivial).
The bosonic and fermionic matter fields are assigned to
spinor representations of $SO(N)$, and are labeled by undotted
and dotted indices $A,B,\ldots$ and $\dot A , \dot B ,\ldots = 1,
\ldots,d$, respectively; the dimension $d$ is thus also the
dimension of the sigma-model target space (which is severely
restricted by supersymmetry). Modulo  higher-order fermionic
terms, the matter Lagrangian can be written as
$$
\lagr^{\rm matter} = \ft14 \sqrt g\,g^{mn} \, P_m^A\,P^A_n -\ft12
i\sqrt g \,\bar{\x}^{\dot A} D\!\slash\, \x^{\dot A} -\ft12  \sqrt g\,
\bar \x^{\dot A} \g^m\g^n\p^I_m\, P^A_n\,\G^I_{A\dot A} +
\cdots\,, \eqno(3)
$$
(our conventions and notation are those of [18,12]).
The derivative $D_m$ acting on the fermions contains the spin
connection $\o_{ma}$ and a connection field $Q_m^{\dot A\dot B}$
associated with the isotropy group $H$ of the coset space.
In contrast to the matter fields, the gravitinos are
inert under $H'$, and therefore only the $SO(N)$ component
of the $H$-connection, $Q_m^{IJ}$, must be included
in the gravitino covariant derivative in (2).
The matrices $\G^I_{A\dot A}$ and their
transpose generate a real (not necessarily irreducible)
representation of the $N$-dimensional Clifford
algebra.\footnote{${}^2$}
{Actually, one must have a representation of the
$(N+1)$-dimensional Clifford algebra in order to encompass
fermion number. Details on three-dimensional supermultiplets will
be published elsewhere. }
The quantities $P^A_m$, whose square constitutes the kinetic term
for the bosons,
are governed by the Cartan-Maurer equations of $G/H$ in the
usual fashion, together with the connections $Q_m^{AB}$ (the $H$
connection acting in the representation appropriate to $P^A_m$).

We are here interested in the reduction of these models to two
dimensions. For the dreibein, we make the standard gauge choice
$$
{e_m}^a = \pmatrix{ {e_\m}^\a  & \rho A_\m   \cr
           \noalign{\vskip 1truemm}
             0  & \rho   \cr}\,,   \eqno(4)
$$
where the lower off-diagonal component has been eliminated by a
local Lorentz ($SO(1,2)$) transformation; we use Greek letters to
denote indices in two dimensions. In two dimensions the
Kaluza-Klein vector $A_\mu$ carries no physical degrees of freedom
and plays the role of an auxiliary field.
The gravitino fields decompose into
two-dimensional gravitino fields $\p^I_\m$ and extra fermion
fields $\P^I$ associated with $\p^I_m$ in the third dimension.

The resulting two-dimensional theory thus contains the zweibein
field ${e_\m}^\a$, the dilaton field $\r$, $N$ gravitino fields
$\p^I_\m$, $N$ extra spinor fields $\P^I$ and the matter fields
incorporated in $P^A_\m$ and the spinors $\x^{\dot A}$. For our
subsequent calculations it is convenient to make the
superconformal gauge choices
$$
e_\m{}^\a = {\rm e}^\s\,\d^\a_\m\,,\qquad \p^I_\mu= i \g_\m
\varphi^I  \,.            \eqno(5)
$$
These gauge conditions require the introduction of the
corresponding ghost and anti-ghost fields:
an anti-commuting vector ghost field $c^\m$, commuting
spinor ghosts $\g^I$, an anti-commuting symmetric
traceless tensor anti-ghost $b^{\m\n}$ and commuting traceless
vector-spinor anti-ghosts
$\b^I_\m$ (so that $b^\m_{\,\m} = \g^\m \b^I_\m =0$). The
vanishing of the corresponding BRST charges on the physical
states effectively imposes the constraint that the stress tensor
associated with the reference metric (cf. (1)) vanishes; the
vanishing of its trace is already guaranteed by the general
argument presented below (1). The conformal factor $\exp \s$ and
the fields $\ve^I$ are well known from conformal field theory,
where they decouple from the physical (transverse) fields by
(super)conformal invariance (at least classically).
The fields $\rho$ and $\P^I$, on the other hand, are the remnants
of the three-dimensional ancestor theory.

It turns out that, after appropriate rescalings of the fermion
fields, the Lagrangian acquires an interesting form in the gauge
(5),
$$
\lagr = \r\big\{\ft12 \pa^2 \s  + \hat\lagr\big\}  \,,\eqno(6)
$$
where $\hat\lagr$ is now independent of the fields $\r$ and $\s$.
Although these fields only play an ancillary role in the
actual calculation of the ultraviolet divergences as they cannot
appear in closed loops, they are crucial for the
final result as we will see. Needless to say,
the rescalings of the various fields are accompanied by appropriate
Jacobians in the functional integral. For this reason it is
premature to conclude that the Lagrangian (6) gives rise to a
delta function, after integrating out the field $\r$; note also
that the moduli space that would be implied by this naive $\r$
integration is infinite. Indeed the result of our calculations
confirms that the generic theory is not trivial in this respect.
In integrating over $\rho$ one should also take into
account the residual (super)conformal transformations preserving the
form of the gauge conditions (5), whose ``volume" must be divided
out of the functional measure. Let us also mention that the form
in which the fields $\rho$ and $\sigma$ appear in (6) suggests
their interpretation as unphysical longitudinal target-space
coordinates [6, 12].

The Lagrangian $\hat\lagr$ contains the contributions from
all fields other than $\r$ and $\s$, including the ghost
fields mentioned above. Suppressing terms quartic in the fermions
and the ghost fields (whose explicit form is
not needed for subsequent calculations), we find
$$\eqalignno{
\hat\lagr &=
\ft14 P^A_\m \,P^{A\m} -\ft12 i\bar\x^{\dot A} \g^\m
(\pa_\m\x^{\dot A} + Q^{\dot A\dot B}_\m \,\x^{\dot B})\cr
&\quad  - i\bar \P^I\g^\m\big(\pa_\m\ve^I +Q^{IJ}_\m\ve^J
-\ft12  \x^{\dot A} \,P_\m^A \,\G^I_{A\dot A}\big) \cr
&\quad +i b^{\m\n}\pa_\m c_\n + \bar\b{}^I_\m \g^\n\g^\m
\big(\pa_\n \g^I + Q^{IJ}_\n \g^J\big) \,. &(7)\cr}
$$

To investigate the short-distance properties of this theory
we employ the standard background field expansion [19,
20], splitting all fields into background and quantum fields.
When expanding
the action, the curvatures $R_{AB}{}^{\!CD}$,
$R_{AB}{}^{\!\dot C\dot D}$ and $R_{AB}{}^{\!IJ}$ appear as
well as their covariant derivatives. For the class of manifolds
that we consider, the (tangent-space) curvatures are $H$-invariant
constants, which are thus covariantly constant with respect to
the $H$-covariant derivatives (i.e. the coset manifold is
symmetric). Furthermore, the group $H$ must
leave the gamma matrices $\G^I$ invariant for reasons of
supersymmetry. This implies the equation
$$
R_{AB}{}^{\! IJ} \, \G^J_{C\dot D}+
 R_{AB}{}^{\! CE} \, \G^I_{E\dot D} +
R_{AB}{}^{\! \dot D\dot E} \, \G^I_{C\dot E}  =0 \,. \eqno(8)
$$
In addition the target space is Einstein, so that the Ricci
tensor satisfies
$$
R_{AB}\equiv R_{AC B}{}^{\!C}= - c\, \d_{AB}\,.          \eqno(9)
$$
Under mild assumptions on the coset decomposition
one can prove that $c=N+\ft 18 d-2$ for $N>4$.\footnote{${}^3$}{
For $N=16$, 12, 10, 9, 8, 6, and 5 supergravity coupled to a
single matter multiplet the coefficient $c$ is just the dual
Coxeter number of the groups $G=E_8$, $E_7$, $E_6$, $F_4$, $SO(8,
1)$, $SU(4,1)$ and  $Sp(2,1)$, which are the (conjectured) target-space
isometry groups for these theories. For $N=8$ supergravity coupled to
$n$ matter multiplets, $G=SO(8,n)$, and $c$ again coincides with
the dual Coxeter number.}
As already mentioned above, these models have not been studied
extensively in the literature, but the above properties can be
verified explicitly for the known theories and are in line
with more general arguments on the structure of generic
three-dimensional supergravity theories with homogenous sigma
models.

{}From (6), it is obvious that the field $\r$ plays the role of a
loop-counting parameter. It is then convenient to
absorb a factor $\r^{1/2}$ into the quantum fields, so that their
kinetic terms appear without a factor $\r$ in front.
We will use dimensional regularization but perform the spinor
algebra in two dimensions so as to preserve supersymmetry. This
should cause no undue harm, as our theory is vector-like and
ambiguities having to do with the definition of $\gamma^3$ do not
arise. Wherever necessary we insert a regulator mass in the
propagators to deal with infra-red divergences.

Let us first discuss the one-loop divergences. Just as for generic
flat-space non-linear sigma models [19], there are no
fermionic loops contributing to infinite one-loop diagrams with
only external bosons. Since, at one loop, the ghost fields
do not contribute either, and since the fields $\r$ and $\s$ cannot
appear in closed loops at all, the calculation here is essentially
the same as for flat-space sigma models. The infinite part of
the one-loop effective bosonic Lagrangian is found to be
$$
\lagr^{(1)}_{\rm DIV} ({\rm bosonic}) = {1\over 2\pi\e}\Big\{\ft12
R_{AB}\, P^A_\m\,P^{B\m} - \ft18 d\, \r^{-2}\,(\pa_\m\r)^2\Big\}\,
.\eqno(10)
$$
At this point, one might be tempted to conclude that the model
is one-loop divergent, because, from (9), the target manifold is
obviously not Ricci-flat, and thus the usual criterion for
one-loop finiteness is not met.
It is here that the fields $\r$ and $\s$
play a role. Because the homogeneous
spaces under consideration are Einstein manifolds (cf. (9)), the first
term in (10) is just the bosonic kinetic term in $\hat\lagr$.
On the other hand, the field equation obtained by varying
$\r$ in (6) tells us that this term is equal to $\pa^2\s$. But this
is a total derivative and can therefore be dropped from (10)! The
second term in (10) can be treated in a similar fashion. Rewriting
it as $\ft18d\{\pa^\m(\r^{-1}\pa_\m\r) - \r^{-1}(\pa^2\r)\}$, we see
that it vanishes by the equation of motion $\pa^2\r = 0$ up to a
total derivative. In summary, all divergences disappear when
the equations of motion are imposed and can thus be absorbed
into divergent redefinitions of the fields $\r$ and $\s$.
In passing we note that this result proves
that the two-dimensional reductions of pure and
Maxwell-Einstein four-dimensional gravity are also one-loop
finite, as these theories lead to $SO(2,1)/SO(2)$ and $SU(2,
1)/(SU(2)\otimes U(1))$ sigma models, whose target spaces are
Einstein manifolds.

Because of the constraints of supersymmetry one expects the
one-loop finiteness to persist for the fermionic terms as well. To
verify this we have also evaluated the infinite terms that are
quadratic in the fermion fields. We record the following terms
$$\eqalignno{
\lagr^{(1)}_{\rm DIV} ({\rm fermionic}) = {1\over
2\pi\e}\Big\{&\ft12 i\bar\x^{\dot A} \g^\m\x^{\dot B}\, P^A_\m\,
D^B \!R_{BA\dot A\dot B}  &(11)    \cr
&+i\bar\P^I \g^\m\ve^J \, P^A_\m\, D^B\!R_{BA}{}^{\!IJ} -\ft12
i\bar\P{}^I \g^\m\x^{\dot A}\, P^B_\m\, \G^I_{A\dot A} \,
R_{AB}\Big\} \,,\cr}
$$
where we made use of the identity (8). As the derivatives on the
curvatures vanish for the class of target spaces that we
consider, we are left with the third term, whose coefficient is
such that the one-loop infinite part of the effective action
takes the form (modulo the ghost fields and terms quartic in the
fermion fields),
$$\eqalignno{
S^{(1)}_{\rm DIV} ={1\over 2\pi\e}\int\di2x\;\Big\{&2c \, {\d
S^{(0)}\over\d\r(x)} -c \,(\r^{-1}\bar\x{}^{\dot A})(x)\, {\d
S^{(0)}\over\d\bar\x{}^{\dot A}(x)}\cr
& -2c\, (\r^{-1}\bar\P{}^I)
(x) \,{\d S^{(0)}\over\d\bar\P{}^I(x)}- \ft14 d\,\r^{-1}(x)\, {\d
S^{(0)}\over\d\s(x)} \Big\} \,.  &(12)\cr}
$$
The result is thus explicitly proportional to the field equations
associated with the classical action $S^{(0)}$. The infinities
can again be absorbed into infinite field redefinitions, and
hence the full theory is one-loop finite.

Let us turn to a discussion of the two-loop divergences in
the bosonic terms of the effective action. First consider the
diagrams with overlapping divergences, which give rise to both
first- and second-order poles in $\e$. It turns out that the
contribution from the ghosts is opposite to
that from the gravitino fields $\P^I$ and $\ve^I$. This
cancellation is consistent with the fact that the  ghost and
gravitino contributions should cancel in the absorptive part of
these diagrams because of unitarity. The single-pole
contributions from the diagrams with overlapping divergences are
proportional to
$$
{1\over 2\pi\e} \r^{-1} \Big[R_{ACDE}\,R_{BC}{}^{\!DE} -R_{AC\dot
D\dot E}\, R_{BC}{}^{\!\dot D\dot E} \Big]\,P^A_\m\,P^{B\m} .
\eqno(13)
$$
The remaining diagrams lead to divergences which,
after removing the subdivergences,
are all proportional to $\e^{-2}$. Having established one-loop
finiteness these terms together with the $\e^{-2}$ contributions
from the diagrams with overlapping divergences
should cancel by virtue of the pole
equations [19]. Therefore (13) represents the only possible
ultra-violet infinities.

The result (13) is similar in form to the corresponding two-loop
result for rigidly supersymmetric sigma models [19], but
there are some important differences. In the absence of
torsion, the fermionic connection (written in
target-space indices) in the rigidly supersymmetric models is
just the Christoffel connection, so that the two
contributions in (13) cancel. However, for locally supersymmetric
models the fermionic connection is in general different.
Nevertheless the expression in (13) can still vanish because the
relevant traces in the dotted and undotted spinor representations
coincide. In the generic coset decomposition that we
used, where the isotropy group equals $SO(N)\otimes H'$,
this is indeed the case, so that these models,
which include the explicitly
known $N=16$ and 8 theories, are two-loop finite.

However, while the arguments for this decomposition
are rather compelling when $N>5$, this is no longer so
for $N\leq4$: for $N=4$, the group
$SO(4)$ is not simple and factors into two $SO(3)$ subgroups, one
acting on the bosons and one on the fermions. Indeed the isotropy
group is reduced and equal to
$H=SO(3)\otimes SO(2)$ (for one matter multiplet). For
$N=2$ the isotropy group equals $SO(2)$, and the explicit
construction of the $N=2$ Lagrangian reveals that bosons and
fermions carry different $SO(2)$ charges 2 and $\ft32$
(these charges are just the helicities of the corresponding
propagating states of $N=1$ supergravity in four dimensions)
[12], so that for
$N=2$ supergravity (13) does {\it not} vanish. To confirm this
conclusion by an independent argument, one may decompose the
relevant representations of $SO(16)$ in the
maximally extended $N=16$ theory with respect to $SU(8) \otimes U(1)$
(corresponding to a decomposition of the $N=16$ multiplets into
$N=2$ multiplets) and verify that the contribution of the $U(1)$
generators does not vanish for $N \leq 4$, indicating that
these models are divergent at two loops.
Incidentally, the purely bosonic theories obtained by dimensional
reduction of gravity in higher dimensions are, of course, not
finite at the two-loop level, as the fermionic contribution is then
absent from (13).

We have thus established two-loop finiteness for a non-trivial
class of interacting field theories. Compared to the standard
supersymmetric sigma models there are many new features related
to local supersymmetry; one of them plays an important role for
the one-loop finiteness. The two-loop finiteness depends,
however, on the details of the symmetric target space, and
therefore on $N$.
Of course the question, which we are unable to answer
at present, is whether the finiteness persists to all orders, and
if so, for which class of theories. Assuming that some of these
theories are finite to all orders, one wonders what the nature
of the critical point could be. Also in this respect our result is
intriguing, as there is only a small number of viable
conformal field theories with extended (local) supersymmetry. Here
it is important to realize that the model
is interacting (even part of the ghost sector is interacting) so
that many of the usual arguments are not always applicable.
Although cosets and algebraic structures play a role in these
models, the standard arguments do not permit to connect them
immediately to conformal models of the (gauged)
Wess-Zumino-Witten-Novikov type.

We emphasize that our results are not directly related to the
two-loop finiteness of supergravity in four dimensions (e.g.
the $N=2$ theory is not
two-loop finite unlike its four-dimensional ancestor!), nor can
any conclusion be drawn from the finiteness in two dimensions for
the corresponding four-dimensional theory. It is clear that the
comparison of short-distance properties of two- and
four-dimensional theories related by dimensional reduction is
subtle. In the reduction to
two dimensions one suppresses infinite towers of massive
Kaluza-Klein states, which contribute to the four-dimensional
short-distance singularities. At the quantum level, the limit of
shrinking the size of the two-dimensional torus to zero (so that
the massive states acquire infinite mass) and the short-distance
limit cannot be interchanged. Furthermore it is not obvious how
to obtain direct information
from the structure of four-dimensional counterterms,
which describe the non-renormalizable sector of the higher-dimensional
theory, especially since
the two-dimensional theory is the result of a variety of
manipulations, such as straightforward reduction, duality
transformations to convert vector fields to scalars and
integrating out auxiliary fields.

\bigskip

\centerline{\bf Acknowledgements}

\noindent
H. Nicolai thanks the CERN Theory Division for hospitality. We
acknowledge valuable discussions with L.~Alvarez-Gaum\'e,
D.~Birmingham and A.H.~Chamseddine.

\medskip

\centerline{\bf References}
\medskip

\item{1.} E.~Braaten, T.L.~Curtwright G.~Ghandour and C.B.~Thorn,
Ann.~Phys. (N.Y.) {\bf 147} (1983) 365,   \hfil\break
J.L.~Gervais and A.~Neveu, Commun.~Math.~Phys. {\bf 100} (1985)
15, \hfil\break
V.G.~Knizhnik, A.M.~Polyakov and A.B.~Zamolodchikov,
Mod.~Phys.~Lett. {\bf A3} (1988) 819, \hfil\break
F.~David, Mod.~Phys.~Lett. {\bf A3} (1988) 1651, \hfil\break
J.~Distler and H.~Kawai, Nucl.~Phys. {\bf B321} (1989) 509.
\item{2.} N.~Seiberg, {\it Notes On Quantum Liouville Theory
and Quantum Gravity}, lectures given at the 1990 Yukawa
International Seminar and at the 1990 Carg\`ese Meeting, preprint
RU-90-29.\hfil\break
E.~D'Hoker, Mod.~Phys.~Lett. {\bf A6} (1991) 745.\hfil\break
L.~Alvarez-Gaum\'e and C.~G\'omez, {\it Topics in Liouville
Theory}, lectures at the 1991 Trieste Spring School, preprint
CERN-TH.6175/91.
\item{3.} S.~Elitzur, A.~Forge and
E.~Rabinovici, {\it Comments on the importance of being
over-constrained}, preprint CERN-TH.6396/92.
\item{4.} D.~Birmingham, M.~Blau, M.~Rakowski and G.~Thompson,
Phys.~Report {\bf 209} (1991) 129.
\item{5.}
C.~Teitelboim, {\it in} ``Quantum Theory of Gravity", ed.
S.~Christensen (Hilger,1984) p.~327, \hfil\break
R.~Jackiw, {\it in} ``Quantum Theory of Gravity", ed.
S.~Christensen (Hilger,1984) p.~403;
Nucl.~Phys. {\bf B252} (1985) 343.
\item{6.} A.H.~Chamseddine, Nucl.~Phys.
{\bf B368} (1992) 98. \hfil\break
T.T.~Burwick and A.H.~Chamseddine, {\it Classical and Quantum
Considerations of Two-dimensional Gravity},
preprint ZU-TH-4/92.
\item{7.} K.~Isler and C.~Trugenberger, Phys.~Rev.~Lett. {\bf 63}
(1989) 834, \hfil\break
A.H.~Chamseddine and D.~Wyler, Phys.~Lett. {\bf B288} (1989) 75.
\item{8.} B.~de~Wit and M.T.~Grisaru, {\it in}
``Quantum Field Theory and Quantum Statistics", Vol.~2, p.~411,
eds. I.A.~Batalin, C.J.~Isham and G.A.~Vilkovisky (Hilger, 1987),
\hfil\break
I.~Halliday, E. Rabinovici, A.~Schwimmer and M.~Chanowitz,
Nucl.~Phys. {\bf B268} (1986) 413.
\item{9.} C.G.~Callan, S.B.~Giddings, J.A.~Harvey and
A.~Strominger, Phys.~Rev. {\bf D45} (1992) 1005.\hfil\break
J.G.~Russo, L.~Susskind and L.~Thorlacius, {\it Black Hole Evaporation
in $1+1$ dimensions}, preprint SU-ITP-92-4
\item{10.} C.~Hoenselaers and W.~Dietz (eds.), ``Solutions of
Einstein's Equations: Techniques and Results" (Springer, 1984).
\item{11.} P.~Breitenlohner and D.~Maison, Ann. Inst. Poincar\'e
{\bf 46} (1987) 215.
\item{12.} H.~Nicolai, {\it Two-dimensional Gravities and
Supergravities as Integrable Systems}, preprint DESY 91-038
(1991), to appear in the proceedings the of 30-th Schladming Winter
School, February 1991.
\item{13.} R.~Geroch, J.~Math.~Phys. {\bf 12} (1971) 918; {\bf
13} (1972) 394.
\item{14.} B.~Julia, {\it in} ``Superspace and Supergravity",
eds. S.W.~Hawking and M.~Ro\v{c}ek (Cambridge Univ. Press, 1980);
{\it in} Johns Hopkins Workshop on Current Problems in Particle
  Physics, ``Unified Field Theories and Beyond" (Johns Hopkins
  University, Baltimore, 1981).
\item{15.} H.~Nicolai, Phys. Lett. {\bf 194B} (1987) 402,
\hfil\break
H.~Nicolai and N.P.~Warner, Commun.~Math.~Phys. {\bf 125} (1989)
384.
\item{16.} D.~Maison, Phys.~Rev.~Lett. {\bf 41} (1978) 521,
\hfil\break
V.A.~Belinskii and V.E.~Sakharov,
Zh.~Eksp.~Teor.~Fiz. {\bf 75} (1978) 1955; {\bf 77} (1979) 3.
\item{17.} E.~Witten, Nucl.~Phys. {\bf B311} (1988/89) 46.
\item{18.} N.~Marcus and J.H.~Schwarz, Nucl.~Phys. {\bf
B228} (1983) 145.
\item{19.} D.~Friedan, Phys.~Rev.~Lett. {\bf 45} (1980)
1057; Ann.~Phys. (N.Y.) {\bf 163} (1985) 318,  \hfil\break
L.~Alvarez-Gaum\'e, D.Z.~Freedman and S.~Mukhi, Ann.~Phys.~(N.Y.)
{\bf 134} (1981) 85.
\item{20.} S.~Mukhi, Nucl.~Phys. {\bf 264} (1986) 640.

\end